\documentclass[prl,aps, twocolumn, showpacs]{revtex4}

\usepackage{graphicx}
\usepackage{epsfig}
\usepackage{amssymb}
\usepackage{epstopdf}
\usepackage{amsmath}
\usepackage{amsthm}
\usepackage{multirow}
\usepackage[usenames]{color}
\definecolor{Red}{rgb}{1,0,0}
\definecolor{Blue}{rgb}{0,0,1}

\begin{document}

\title{Observation of the ground-state-geometric phase in a Heisenberg XY model}
\author{Xinhua Peng$^{1}$}
\email{xhpeng@ustc.edu.cn}
\author{Sanfeng Wu$^{1}$}
\author{Jun Li$^{1}$}
\author{Dieter Suter$^{2}$}
\email{Dieter.Suter@tu-dortmund.de}
\author{Jiangfeng Du$^{1}$}
\email{djf@ustc.edu.cn}

\affiliation{$^{1}$Hefei National Laboratory for Physical Sciences at Microscale and
Department of Modern Physics, University of Science and Technology of China, Hefei, Anhui 230026, People's Republic of China}
\affiliation{$^{2}$Fakult\"{a}t Physik, Technische Universit\"{a}t Dortmund, 44221 Dortmund, Germany}

\begin{abstract}
Geometric phases play a central role in a variety of quantum
phenomena, especially in condensed matter physics.
Recently, it was shown that this fundamental concept exhibits a connection to quantum phase
transitions where the system undergoes a qualitative change in the ground state
when a control parameter in its Hamiltonian is varied.
Here we report the first experimental study using the geometric phase as a topological test of quantum transitions
of the ground state in a Heisenberg XY spin model.
Using NMR interferometry, we measure the geometric phase for different adiabatic circuits that do not
pass through points of degeneracy.
\end{abstract}
\maketitle

When a quantum system is subjected to a cyclic adiabatic
evolution, it returns to its original state but may acquire a geometric
phase factor in addition to the dynamical one.
Berry made this surprising discovery in 1984 \cite{berry392}, so that this is also
known  as Berry's phase.
Later this phase was generalized in various directions, to include a more general case of noncyclic and
nonadiabatic evolution \cite{AA87}, and even the case of mixed states.
Geometric phases (GP) have been observed in a wide variety of physical systems, e.g., in
spin-polarized neutrons \cite{neutron59}, nuclear magnetic resonance (NMR) \cite{NMR60} and
superconducting systems \cite{solid318}.
Moreover, GP has found
applications to many areas, such as molecular dynamics, many-body
systems and quantum computation \cite{GPQC00,GPphysics1}.

Very recently, the GP of many-body systems has been shown to be
closely connected to quantum phase transitions (QPTs), an important
phenomenon in condensed matter physics \cite{Carollo05,slzhu05}.
QPTs occur at zero temperature and describe abrupt changes in the
properties of the ground state resulting from the presence of level
crossings or avoided crossings \cite{Sachdev99}.
Recently, different methods related to quantum information have been
developed for characterizing QPTs, including the fidelity
\cite{overlap06}, quantum entanglement \cite{entangle02,penPRL} and some
other geometric properties \cite{loschmidt06}.
The GP, which is a measure of the curvature of Hilbert space, can
reflect the energy level structure to fingerprint certain features of QPTs.
Carollo and Pachos \cite{Carollo05} demonstrated that the
GP difference between the ground state and the first excited state
encounters a singularity  when the system undergoes a QPT in the XY
spin chain. Zhu \cite{slzhu05} revealed that GP associated with its
ground state exhibits universality, or scaling behavior, around the
critical point. Besides the study in the thermodynamical limit, it
was also shown that the GP could be used to detect level crossings
for a two-qubit system with XY interaction \cite{Sangchul09}.
As a complement to these theoretical investigations, it appears highly desirable
to have experimental evidence for these effects.

In this Letter, we report the first experiment that shows this
important connection between the GP and the energy level structure
(i.e., level crossing points) in a Heisenberg XY spin model.
In our experiment, the system Hamiltonian changes adiabatically along a closed trajectory
in parameter space while the system, which is in the ground state of the Hamiltonian,
accumulates a GP.
Depending on the region in parameter space, the resulting GP is zero or has a finite value.
These regions in parameter space are separated by a line where the ground state of the system
becomes degenerate \cite{Sangchul09}.
Using adiabatic state preparation and NMR interferometry, we
observe the transitions of GP on both sides of the level crossings point.
This experiment might be viewed as a first meaningful step to use GP as a fingerprint for ovserving QPTs.

Consider a one-dimensional spin-1/2 XY model in a uniform external magnetic field along the $z$ axis:
\begin{eqnarray}
\mathcal{H}(\lambda,\gamma)&=&-\sum_{j}(\frac{1+\gamma}{2}\sigma^{j}_{x}\sigma^{j+1}_{x}
+\frac{1-\gamma}{2}\sigma^{j}_{y}\sigma^{j+1}_{y})-\frac{\lambda}{2}\sum_{j}\sigma^{j}_{z}, \nonumber
\end{eqnarray}
where $\sigma_{\nu}^{k} (\nu=x,y,z)$ denote the Pauli matrices for qubit $k$, $\lambda$ is the
strength of the external magnetic field, and $\gamma$ measures the anisotropy of the coupling strength in the XY plane.
This model is exactly solvable and can be diagonalized by the Jordan-Wigner
transformation, Fourier transformation and then Bogoliubov transformation \cite{diagXY}.
However, it still contains a rich phase structure \cite{Sachdev99}.
Barouch and McCoy \cite{PRA71}
investigated the statistical mechanics of this model in the
thermodynamical limit and showed that a circle ($\lambda^2 + \gamma ^2 =1$) separates the oscillatory phase (inside) from the para- or
ferro-magnetic phase (outside).
At the level crossing or avoided crossing between ground state and
first excited state, the ground state changes discontinuously.
As a result, the GP associated with the ground state also changes discontinuously.
Theoretical work has demonstrated the close relation
between GPs and the energy level structures, thereby revealing the
ground-state properties \cite{Carollo05, slzhu05}, even in the
two-qubit case \cite{Sangchul09}.

We now consider the GP that results in this system if the Hamiltonian rotates around the $z$-axis,
$ \tilde{\mathcal{H}}(\lambda,\gamma,\phi) =
U_{z}^{\dagger}(\phi) \mathcal{H}(\lambda,\gamma)U_{z}(\phi) $
with $U_{z}(\phi)=\prod_k e^{-i\frac{\phi}{2}\sigma_{z}^{k}}$ \cite{Carollo05}.
$\tilde{\mathcal{H}}$ has the same spectrum as $\mathcal{H}$,
independent of $\phi$.
Here we study a minimal model of two qubits
coupled by an XY-type interaction \cite{Sangchul09}.
The eigenvalues of $\tilde{\mathcal{H}}$ are $\pm1$ and $\pm r$,
where $r = \sqrt{\lambda^2+\gamma^2}$.
The ground state is
\begin{eqnarray}
|\Psi_{g} (\phi) \rangle &=& \left\{
\begin{aligned}
 \frac{1}{\sqrt{2}}(|01\rangle + |10\rangle), & & &  r <  1\\
\cos\frac{\theta}{2}|00\rangle + \sin\frac{\theta}{2}e^{-i2\phi}|11\rangle, & & & r > 1
\end{aligned}
\right.
\end{eqnarray}
where $\tan\theta=\gamma/\lambda$.
For $r < 1$, the ground state is thus invariant; for $r=1$, it is doubly degenerate;
and for $r > 1$, it is spanned by the two states $|00\rangle$ and $|11\rangle$,
with coefficients that depend on the angle $\theta$.

\begin{figure}
  \includegraphics[width=0.99\columnwidth]{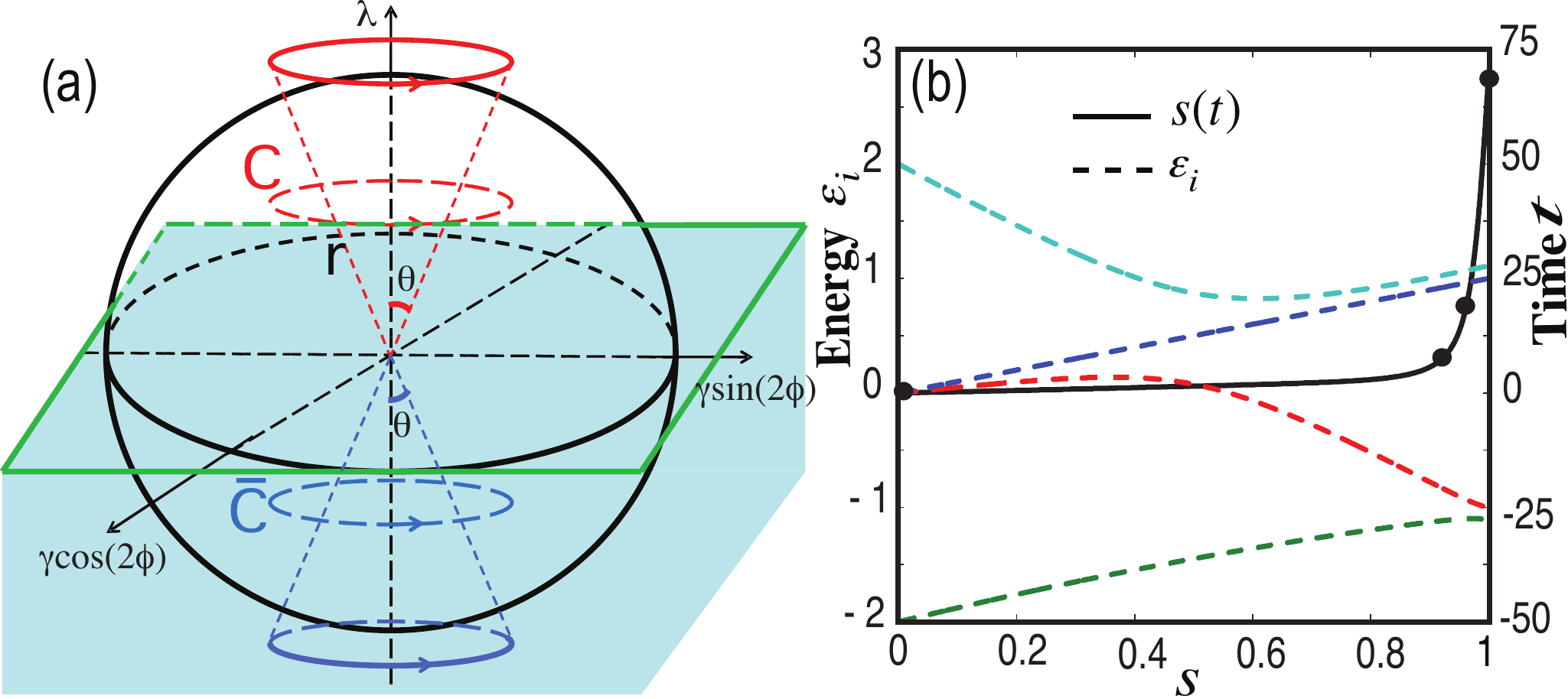}\\
  \caption{(Color online)
  (a) Parameter-space representation of the cyclic adiabatic evolutions that generate the GP.
 Two closed paths $C$ and $\bar{C}$, related by inversion symmetry, were combined for observing a purely GP.
 The cycles are horizonal, i.e., $\lambda$ is constant and $\gamma$ is constant.
 The observed GP depends on the angle $\theta$ if the circles are outside the sphere $r=1$ (shown in black)
 and vanish if the curves are inside the sphere. 
 (b) Energy level diagram of the time-dependent $\mathcal{H}_{ad}(s)$ for ASP (denoted by the dashed lines), and the optimal function of
  adiabatic parameter $s(t)$ (denoted by the solid line) calculated for a constant adiabaticity factor, when $\gamma=0.5$, $\lambda=0.9920$.
  The black dots represent the experimental values for the discretized scan. }
  \label{geophase}
\end{figure}

If we let the Hamiltonian travel along a cyclic path in the parameter space
$(\lambda, \gamma, \phi)$, we can consider the subspace spanned by $|00\rangle$ and $|11\rangle$,
which contains the ground state, as a pseudo-spin 1/2, where the spin
evolves in an effective magnetic field
$\mathbf{B} = r (\sin \theta \cos 2\phi, \sin \theta \sin 2\phi, \cos \theta)$.
Using the standard formula $\beta_{g}= i \oint_{0}^{\pi} \langle \Psi_g \vert
\partial_{\phi}\vert \Psi_g \rangle$ \cite{berry392}, the ground state accumulates a GP
\begin{eqnarray}
\beta_{g}(\phi:0\rightarrow\pi) &=& \left\{
\begin{aligned}
 0, & & &  r < 1,\\
 \pi(1 - \cos\theta),& & & r > 1.
\end{aligned}
\right.
\end{eqnarray}
As shown in Fig.~\ref{geophase}(a), it is useful to represent the trajectory in a parameter space spanned by
$\gamma \cos(2 \phi)$, $\gamma \sin(2 \phi)$ and $\lambda$.
Here, the sphere with radius $r=1$ marks the points where the Hamiltonian is degenerate.
Inside this sphere ($r<1$), the GP vanishes, while it has a finite value that depends on the opening angle $\theta$
of the cone subtended by the circuit.
A special case is the XX spin model (i.e., $\gamma = 0$).
Here, the GP always vanishes, because the operation $U_{z}$ does not change
the Hamiltonian of the system.
While we are considering here only a minimal two-spin model, the ground state and the ground state energy
of the XY model in the thermodynamic limit are similar \cite{PRA71}.

When the system undergoes the cyclic adiabatic evolution along
$\tilde {\mathcal{H}}$, there will also be an additional dynamic
phase generated, relative to the instantaneous energy of the system,
besides the GP. Hence, in order to acquire the pure geometric part,
we have to eliminate the dynamical contribution. To eliminate the
dynamical contribution to the phase shift, we combine two
experiments with the closed paths $C$ and $\bar{C}$ \cite{AA87}, which generate the
same geometrical phases, but opposite dynamical phases. The two
trajectories have the same geometrical shape (cones), but their
Hamiltonians $\tilde {\mathcal{H}}$ and $- \tilde {\mathcal{H}}$ and
thus their dynamical phases add to zero. 
During the first period,
the Hamiltonian $\tilde{\mathcal{H}} (\lambda, \gamma,\phi)$ follows
the closed curve $C$ in the parameter space $\mathbf{r} = (r,\theta,\phi)$,
with $\phi$ changing from 0 to $\pi$, as schematically shown by the
the red circles (labeled by $C$ in the upper part) for $\lambda > 0$ in
FIG. \ref{geophase}(a). 
During the second period, the Hamiltonian $
- \tilde {\mathcal{H}} = R_{kz}^{\dagger}(\pi) \tilde{\mathcal{H}}
(-\lambda, \gamma,\phi)R_{kz}(\pi)$ follows the curve $\bar{C}$,
shown in the lower part of FIG. \ref{geophase}(a). Here $R_{kz}(\pi)
= e^{-i\frac{\pi}{2}\sigma_{z}^{k}}, (k = 1 \mbox{ or } 2)$ rotates
one of the two spins around the z-axis. For the circuit $C$, the
resulting phase is $\beta_C = \pi(1-\cos \theta) - r T$, where $T$
is the cycle time, where we have assumed $r>1$. For $\bar{C}$,
$\beta_{\bar{C}} = \pi(1-\cos \theta) + r T$ because the sign of the eigenvalue of the state $\vert \Psi_g \rangle$ changes for $
- \tilde {\mathcal{H}}$. The sum of the two
phases, $\beta_C+\beta_{\bar{C}} = 2 \pi(1-\cos \theta)$ is thus
purely geometrical. If $r<1$, the dynamical component changes to $ -T$ for $C$ and $T$ for $\bar{C}$ while the GP vanishes. 

\begin{figure}
  \includegraphics[width=0.9\columnwidth]{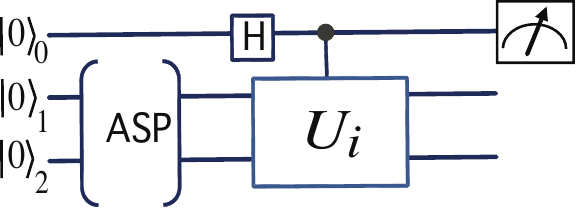}\\
  \caption{Quantum circuit for measuring the ground-state GP.
  $H$ is the Hadamard gate, and following the adiabatic state preparation (ASP), the operation $U_i$ performs a cyclic
  adiabatic evolution of the system qubits (1 and 2) conditionally when the ancilla path qubit $0$ is in the state $\vert 1 \rangle$.}
\label{Uadi}
\end{figure}

For measuring the GP, we use NMR interferometry \cite{NMR60,pengPRANMR}.
This requires an ancilla qubit that is coupled to the system undergoing the circuit.
FIG. \ref{Uadi} shows schematically the experiment, including the adiabatic state preparation (ASP)
of the two qubit system into the ground state of the Heisenberg XY model,
and the generation of a superposition of the ancilla qubit by a Hadmard gate.
The subsequent adiabatic circuit $U_i$, which is conditional on the state of the ancilla qubit,
implements the interferometer
$
\mathcal{U}_i = \vert 0 \rangle \langle 0 \vert_a \otimes \mathbf{1} +  \vert 1 \rangle \langle 1 \vert_a \otimes U_i,
$
where $\mathbf{1}$ represents a $4\times 4$ unit operator and the unitary operator $U_i$
is the cyclic adiabatic evolution on the system qubits along the chosen path $C$ or $\bar{C}$.
The phase acquired during this path appears then directly as a relative phase in the superposition of the two ancilla states
and can be measured in the NMR spectrum of the ancilla spin.

The experiment was carried out on a Bruker Avance III 400MHz (9.4 T)
spectrometer at the room temperature.
The three qubits $0$, $1$ and $2$ in the
quantum circuit (FIG. \ref{Uadi}(a)) were represented
by the $^{1}$H, $^{13}$C, and $^{19}$F nuclear spins in
Diethyl-fluoromalonate.
The relaxation times for all three spins are $T_{2} \approx 1s$.
The natural Hamiltonian of this system is
$\mathcal{H}_{NMR} =
-\sum_{i=0}^2 \frac{\omega_{i}}{2}\sigma_{z}^{i} +
\sum_{i<j} \frac{\pi J_{ij}}{2}\sigma_{z}^{i}\sigma_{z}^{j}$,
where $\omega_{i}$ is the Larmor frequency for spin $i$ and
$J_{ij}$ are the coupling constants $J_{01} = 160.7$ Hz, $J_{12} =
-194.4$ Hz and $J_{02} = 47.6$ Hz. As the sample is not labeled, the
relative phase information on $^{1}H$ at the end of the quantum
circuit was obtained through the $^{13}$C spectrum by a SWAP
operation between $^{13}C$ and $^{1}H$ \cite{penPRL}.

In the experiment, we first initialized the system into the pseudopure state (PPS) $\rho_{000}
=\frac{1 - \epsilon}{8} \mathbf{1}+ \epsilon|000 \rangle \langle 000 |$ by spatial
averaging \cite{penPRL}, with the polarization $\epsilon \approx 10^{-5}$.
Then we prepared the ground state of the Heisenberg XY Hamiltonian by an adiabatic passage:
A rf pulse rotated the spins from the $z$- to the $-x$-axis, i.e., to the ground state of
$\mathcal{H}_0=\sum_{i}\sigma_{x}^{i}$, and then
this Hamiltonian was slowly changed into the target XY Hamiltonian $\mathcal{H}(\lambda,\gamma)$,
always fulfilling the adiabatic condition
$\kappa \ll 1$ \cite{QM}.
This assures that the resulting final state is close to the desired ground state of the XY model.
We optimized the time dependence of the transfer by choosing
$\mathcal{H}_{ad}(t)=[1-s(t)]\mathcal{H}_{0}+s(t) \mathcal{H}$ with $0 \le s(t) \le 1$. 
The solid line in FIG. \ref{geophase}(b) shows the corresponding time dependence for a constant $\kappa$.
The time dependence of $s(t)$ was chosen such that the adiabaticity parameter $\kappa < 0.25$ at all times.

In the experiment, the adiabatic transfer was performed in discrete steps. 
The parameter $s(t)$ therefore assumes discrete values $s_m$ with $m = 0,...,M_P$,
and for each period of duration $\delta$, the corresponding Hamiltonian
$\mathcal{H}_{ad}[s_m]$ was generated by a multiple pulse sequence:
$U_{P}(\delta)=e^{-i\mathcal{H}_{ad}[s_m]\delta}
= e^{-i[1-s_m] \mathcal{H}_{0}\frac{\delta}{2}} e^{-is_m \mathcal {H}(\lambda,
\gamma)\delta} e^{-i[1-s_m] \mathcal {H}_{0}\frac{\delta}{2}} + O(\delta^3)$, via the use of Trotter's formula. 
$\delta$ and $M_P$ were chosen by simultaneously considering this stepwise approximation and the adiabaticity criterion. 
The experimental values $s_m$
for the discretized scan are represented by black dots in FIG. 1(b).
The theoretical fidelity of this stepwise transfer process was $>0.99$,
and the experimental fidelity was $>0.98$.

After the preparation of the ground state, we applied the cyclic adiabatic variation $C$ or $\bar{C}$.
The corresponding control operation
$\mathcal{U}_{C}$ or $\mathcal{U}_{\bar{C}}$
was generated in the form of a discretized adiabatic scan, as described for the ASP part.
Again, the parameters of the scan were optimized to keep the fidelity$ >0.99$.
At the end of the scan, the accumulated phase was measured.

\begin{figure}
  \includegraphics[width=0.8\columnwidth]{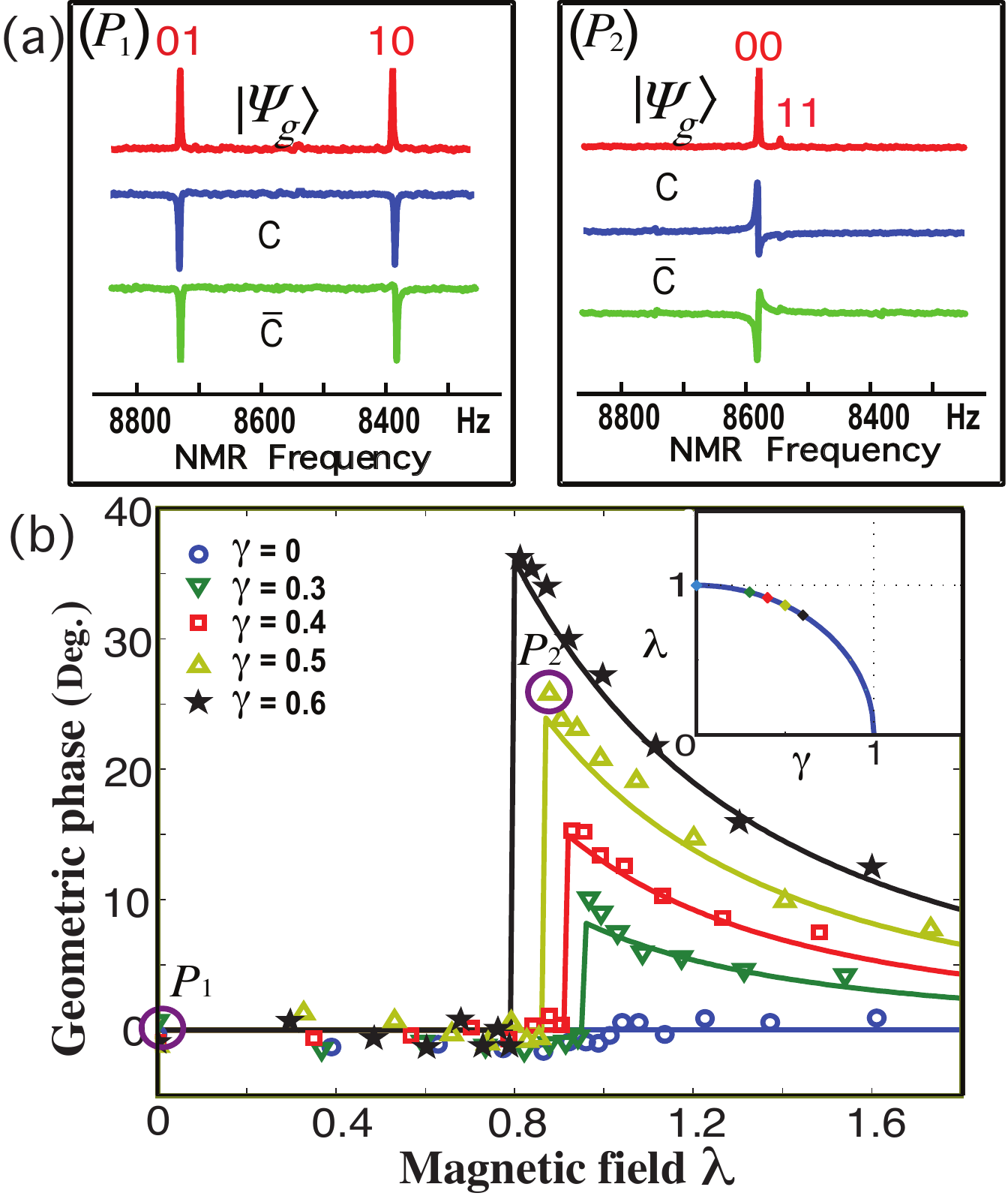}\\
  \caption{(a) Experimental NMR spectra for two specific parameter sets, $P_1$ for $\mathcal{H}(0, 0.5)$ and $P_2$ for $\mathcal{H}(0.878, 0.5)$. From top to
bottom, the spectrum corresponds to the initial ground state
$\vert \Psi_g(\lambda,\gamma)\rangle$, $U_C\vert \Psi_g(\lambda,\gamma)\rangle$ for the
adiabatic path $C$ and $U_{\bar{C}}\vert \Psi_g(\lambda,\gamma)\rangle$ for the
adiabatic path $\bar{C}$.
(b) Measured ground-state GP of
the Heisenberg XY model (points) for different parameter sets
($\lambda, \gamma$) compared to the theoretical expectations (solid curves).} \label{expresult}
\end{figure}

Fig. \ref{expresult} (a) shows two
representative examples of the resulting data: The spectra on the
left hand side correspond to the states before the adiabatic circuit, after traversing
the circuit $C$, and after traversing $\bar{C}$ for the Hamiltonian parameters $(\lambda
,\gamma) = (0,0.5)$. Clearly, in this
situation, we are within the sphere $r=1$, so the ground state of
the system is $\frac{1}{\sqrt{2}}(|01\rangle+|10\rangle)$. This is
verified by the experimental data, where only the two resonance
lines are visible that correspond to the states $|01\rangle$ and
$|10\rangle$ of the system. In the initial state, the two lines
appear in absorption; this corresponds to the reference phase
$\varphi = 0$. During the circuit $C$ or $\bar{C}$, which is
traversed over a time $T=3$, the system should acquire a phase $T$.
In the experimental data, we find that the lines are inverted; a
numerical analysis of the lines yields phases of $\beta_t(C)
\approx 170.6^{\circ}$ and $\beta_t(\bar{C}) \approx -173.0^{\circ}$. Thus the resulting GP $\beta_g = (\beta_t(C) +
\beta_t(\bar{C}))/2 \approx -1.2^{\circ}$, which is close to the
theoretically expected value of zero. The right-hand part of Fig. \ref{expresult}(a) shows the corresponding
results for an adiabatic circuit outside of the sphere $r=1$, where
we expect to observe a non-vanishing GP. In this case, we
observe clearly different phases for the two circuits, whose
duration is now $T=10.7$. The measured phases are  $\beta_t(C) \approx -
62.6^{\circ}$ and $\beta_t(\bar{C}) \approx 114.2^{\circ}$, corresponding
to a GP of $25.8^{\circ}$.

Fig. \ref{expresult} (b) shows the GP
measured for different parameters  ($\lambda, \gamma$). The symbols
show experimental data points, while the curves that connect the
points show the theoretically expected GP as a function of the
magnetic field strength $\lambda$, for a constant anisotropy
parameter $\gamma$. In all cases, the observed GPs are compatible
with the theoretically expected values: zero if the parameters
($\lambda, \gamma, \varphi$) fall inside the sphere with radius
$r=1$, a sudden increase to the maximum value just outside the
sphere, where the opening angle $\theta$ of the cone subtended by
the circuit reaches a maximum, and then decreasing as the circuit
$C$ is moved away from the origin. Increasing values of $\gamma$
correspond to larger circles $C$ and thus bigger values of $\theta$.
The points marked $P_1$ and $P_2$ correspond to the spectra shown in
the upper part of the figure.

The relevant sources of experimental errors mainly came from
undesired transitions induced by the time-dependent Hamiltonian,
inhomogeneities of rf fields and static magnetic fields, imperfect
calibration of rotations and relaxation.
We used a numerical optimization procedure to minimize undesired transitions during the adiabatic passage.
The durations of individual experiments ranged from $30$ ms to $90$ ms, short compared to the relaxation time $T_2 \sim 1s$.
The experimental error of the geometric phase was less than $3^\circ$.
The imperfection of the initial state would also contribute to this.
Using the experimentally reconstructed density matrices for the initial states,
we found that this effect contributed $\approx 1^\circ$ to the errors.

In summary, we have detected the ground-state GP in the Heisenberg XY model,
after preparing the initial state by an adiabatic passage.
The Heisenberg XY model was simulated by a multiple-pulse
sequence, and the phase was measured by NMR interferometry.
Our proof-of-principle experiment illustrates that the ground-state GP
can serve as a fingerprint of the energy-level crossing points that result in a QPT in the thermodynamic limit.
The ground-state GP is a robust indicator that is immune to some experimental imperfections \cite{Robustness} and provides an experimental method
that does not need to cross the critical point.

It would be very interesting to extend this experiment to larger spin systems.
For this, two issues are relevant: (\textit{i}) the
effectiveness of the ASP and (\textit{ii}) the realization of quantum
circuit consisting of a quantum interferometer and quantum simulation.
For the first issue, although a decisive mathematical analysis of
the efficiency of ASP is difficult, numerical simulations (up
to 128 qubits) \cite{ASPsim} indicate a polynomial growth of
the median runtime of an adiabatic evolution with the system size.
On the second issue, quantum interferometry has become
a mature technique, and the Heisenberg XY model has
been efficiently simulated by a universal quantum circuit only
involving the realizable single- and two-qubit logic gates \cite{XYcircuit09}.
Moreover, the diagonalization theory of the XY model
shows a valid energy gap between the two lowest energies, which
guarantees the viability of the cyclic adiabatic evolution to
generate the ground-state GP, even in the thermodynamic limit
\cite{Carollo05}.
Recent research also shows that a 10-qubit
system already represents a good approximation to the
thermodynamical limit \cite{XXEPJ}.
Therefore, the present scheme is
in principle applicable to larger spin systems, when the
technical difficulties in building a medium-scale
quantum computer are overcome. This significant connection between GPs and QPTs is not a specific feature of the XY model, but remains valid in a general case  \cite{Carollo05,slzhu05}.
We hope that this experimental work will contribute to an improved understanding of
the ground-state properties and QPTs in many-body quantum systems.

We thank S. L. Zhu for helpful discussion. This work was supported by NNSFC, the CAS and NFRP,
and by the DFG through Su 192/19-1.

\section{Supporting Online Material}

\subsection{Quantum simulator and characterization}

\begin{figure}
  \includegraphics[width=0.99\columnwidth]{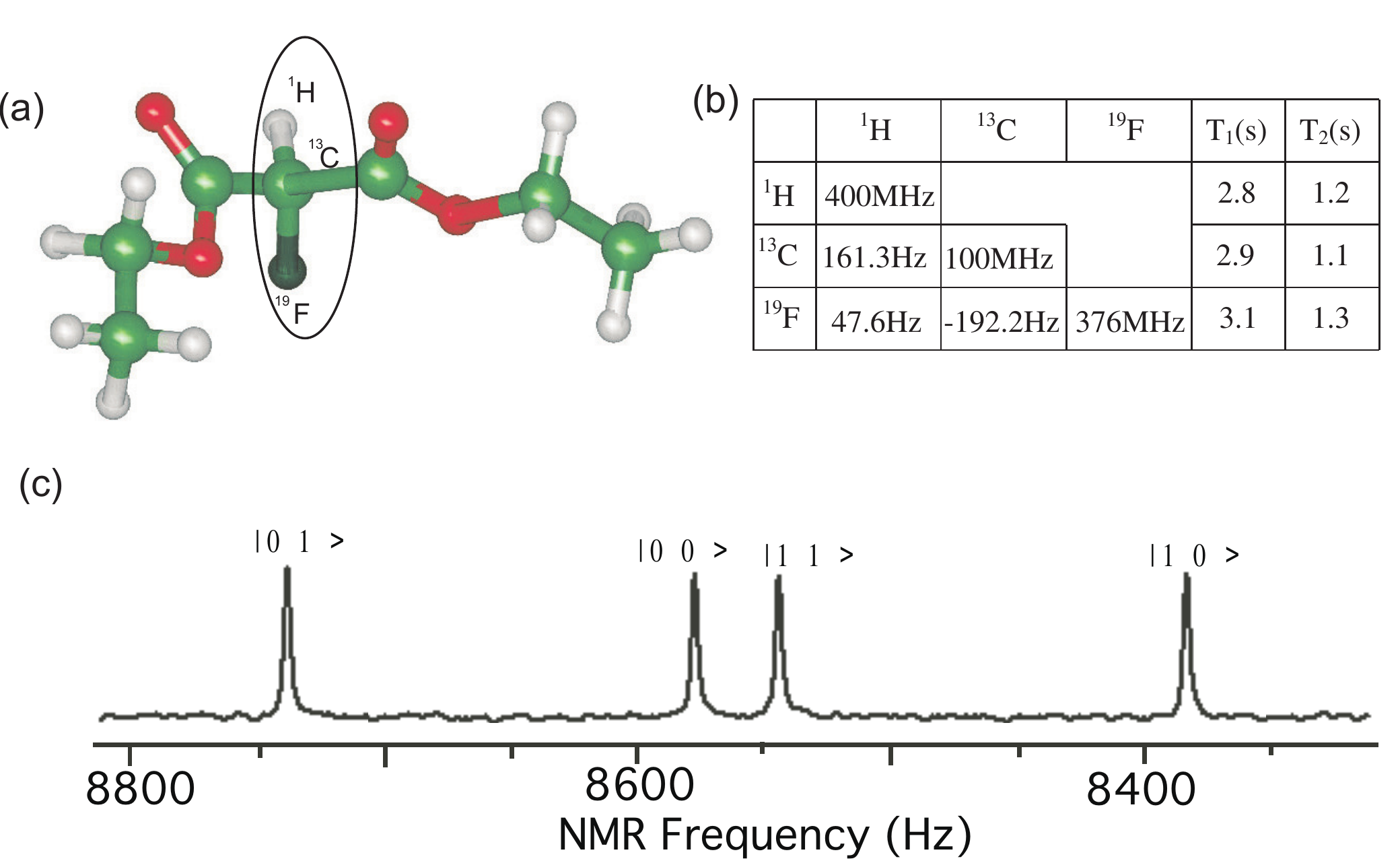}\\
  \caption{(Color online) Relevant properties of the quantum register. (a)
The molecular structure of Diethylfluoromalonate
and the oval marks the three spins used as qubits. (b) The relevant NMR parameters: the resonance frequencies $\omega_i$ (on the diagonal), the $J$-coupling constants $J_{ij}$ (below the diagonal), and the relaxation times $T_1$ and $T_2$
in the last two columns. (c) NMR spectrum of the $^{13}$C obtained through a
read-out $\pi/2$ pulse on the equilibrium state, where the four resonance
lines are labeled by the corresponding states of the two other qubits.}
  \label{sample}
\end{figure}

For the quantum register for these experiments, we selected the $^{1}$H, $^{13}$C, and
$^{19}$F nuclear spins of Diethyl-fluoromalonate dissolved in d-chloroform. 
The relevant system parameters are listed in Fig. \ref{sample}. 
Experiments were carried out at room temperature, using a
Bruker Avance III 400 MHz (9.4 T) spectrometer equipped
with a QXI probe with pulsed field gradient.

\begin{figure}
  \includegraphics[width=0.8\columnwidth]{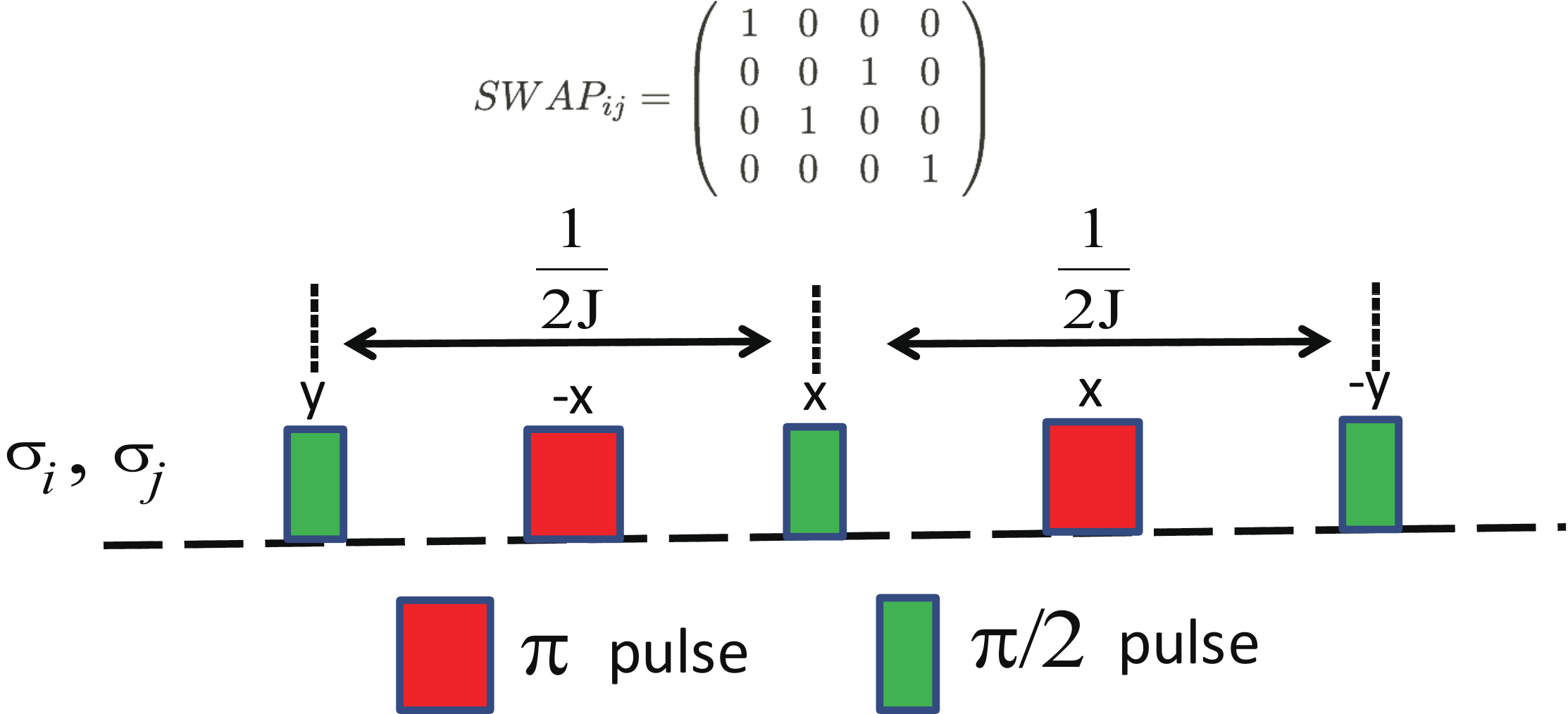}\\
  \caption{Pulse sequence for implementing the SWAP operation on spins $\sigma_i$ and $\sigma_j$ with the
corresponding matrix representative. Narrow rectangles
indicate $\pi/2$ pulses and wide rectangles $\pi$ pulses, and the pulse phases are indicated above them.}
  \label{swap}
\end{figure}

Because we used an unlabeled sample, the molecules with a $^{13}$C nucleus, which we used as the quantum register,
were present at a concentration of about 1\%.
The $^{1}$H and $^{19}$F spectra were dominated by signals from the 2-qubit molecules
containing the $^{12}$C isotope, while the signals from the quantum register with the $^{13}$C nucleus appeared only as small (Å0.5\%) satellites. 
To effectively separate this signal from that of the
dominant background,  transfered the
state of the $^{1}$H and $^{19}$F qubits to the $^{13}$C qubit by a SWAP gate and read the state
through the $^{13}$C spectrum. 
The matrix representation of the SWAP operation  for two spins $\sigma_i$ and $\sigma_j$ is shown in Fig. \ref{swap}, 
together with the corresponding pulse sequence.

\subsection{Experimental procedure}

The experiment, summarized in Fig. 2(a) in the paper, includes three steps: ($i$) Adiabatic state preparation (ASP): to prepare the ground state of the XY spin model $\mathcal{H}(\lambda, \gamma)$ at different points of the Hilbert parameter space $(\lambda, \gamma)$ by adiabatic evolution; ($ii$) NMR interferometry: to generate the pure geometric phase (GP) on one of the two paths where an auxiliary spin is introduced; ($iii$) Phase measurement: to obtain the GP by measuring the relative phase between the two paths via quadrature detection in NMR.

\subsection{1. Adiabatic state preparation}

To prepare the system in the ground state of a Hamiltonian $\mathcal{H}(\lambda, \gamma)$, 
the system can start with an initial Hamiltonian $\mathcal{H}(0) = \mathcal{H}_0$, whose ground state $\vert\psi_g(0)\rangle$ is known and then it is adiabatically driven to the target Hamiltonian  $\mathcal{H}(T) = \mathcal{H}(\lambda, \gamma)$. 
A time-dependent Hamiltonian $\mathcal{H}_{ad}(t)$ smoothly interpolates between $\mathcal{H}(0)$ and $\mathcal{H}(T)$:
\begin{equation}
\mathcal{H}_{ad}(t)=[1-s(t)]\mathcal{H}_{0}+s(t)\mathcal{H}(\lambda, \gamma) \,, \label{e.HSys}
\end{equation}
where the function $s(t)$ varies from 0 to 1 to parametrize the interpolation.
If the quantum system starts in $\vert \psi_g(0) \rangle$ and the variation of $\mathcal{H}(t)$ is adiabatic,
the final state reached will be close to the ground state of $\mathcal{H}(T) = \mathcal{H}(\lambda, \gamma)$.
To ensure that the system is prepared in the ground state of $\mathcal{H}(\lambda, \gamma)$,
the sufficiently slow variation of $\mathcal{H}_{ad}(t)$ means that the traditional adiabatic condition \cite{Messiah:1976aa}
\begin{equation}
\left|  \frac{\langle \psi _{g}(t)\left| \dot{\psi}_{1e}(t)\right\rangle}{\varepsilon _{1e}(t)-\varepsilon _{g}(t)} \right|  \ll 1
 \label{AC}
\end{equation}
is fulfilled, where $\vert \psi _{g}(t) \rangle$ and $ \vert \psi_{1e}(t) \rangle$ refer to the instantaneous ground state
and the first excited state, respectively, and $\varepsilon _{g}(t)$,
$\varepsilon _{1e}(t)$ are the corresponding energies.

To do this, we chose the initial Hamiltonian of the system $\mathcal{H}_{0} =\sigma_{x}^{2}+\sigma_{x}^{3}$, whose ground state  $\vert\psi_g(0)\rangle _{23} = \frac{1}{\sqrt{2}} (\vert 0 \rangle - \vert 1 \rangle) \otimes  \frac{1}{\sqrt{2}} (\vert 0 \rangle - \vert 1 \rangle)$ is well-known. Starting from the thermal equilibrium state, the system was first initialized into a pseudopure state (PPS) $\rho_{000}
=\frac{1 - \epsilon}{8} \mathbf{1}+ \epsilon|000 \rangle \langle 000 |$ by spatial
averaging \cite{PPS}, where $\mathbf{1}$ represents the unity operator and
$\epsilon \approx 10^{-5}$ represents the thermal polarization. The $\vert 0 \rangle$ and $\vert 1 \rangle$ states
correspond to the two eigenstates of $\sigma_z$ Ñ the spin-up and
spin-down states, respectively. The normalized deviation density matrix of the
PPS \cite{PPS}, $\rho_{\Delta} \equiv [\rho  - (1 - \epsilon) \mathbf{1}/8] / \epsilon $ was reconstructed by quantum state tomography
\cite{PPS,QST}, which involves the application of 7 readout pulses and recording of the spectra of all three channels to obtain
the coefficients for the 64 operators comprising a complete operator basis of the three-spin system. The experimentally determined state fidelity was $F=\frac{Tr(\rho^{000}_{exp}*\rho_{th})}{\sqrt{Tr((\rho^{000}_{exp})^2)*Tr(\rho_{th}^2)}} \approx 0.99$. 
Then the initial state $\vert 0 \rangle_1 \otimes \vert\psi_g(0)\rangle _{23}$ was prepared by two pseudo-inverse-Hamdamard gates, 
i.e., $ [\pi /2]_{-y }$ pulses on the system qubits 2 and 3.

An optimal function of $s(t)$ determines the efficiency of ASP. To find the optimal interpolation function $s(t)$ for the adiabatic process, we rewrite the adiabatic condition of Eq. (\ref{AC}) as
\begin{equation}
 \left| \frac{d s(t)}{dt} \right| \ll \frac{| \varepsilon _{1e}(t)-\varepsilon _{g}(t) |^2}{ \left| \langle \psi _{g}(t) \vert  \frac{ \partial \mathcal{H}_{ad}(s)}{\partial s} \vert \psi_{1e}(t) \rangle \right|} = \chi,
 \label{AC2}
\end{equation}
which defines the optimal sweep of the control parameter $s(t)$ with the scan speed $\frac{d s(t)}{dt}$.
The required time dependence of $s(t)$ was numerically optimized for constant adiabaticity parameter $\kappa= \frac{d s(t)}{dt}/\chi$, 
represented by the solid line in Fig. 1 (b) in the paper for the Hamiltonian  $\mathcal{H}(\lambda, \gamma) = \mathcal{H}(0.992, 0.5)$.
The time dependence of $s(t)$ was chosen such that the adiabaticity parameter $\kappa < 0.25$ at all times.

For the experimental implementation,  we have to discretize the continuous adiabatic passage
into $M_P+1$ segments (i.e., $s_m = s(\frac{m}{M_P}T_P) $ with $m = 0,...,M_P$, $s_0 = 0$ and $s_{M_P} = 1$), 
and generate the instantaneous discretized Hamiltonian $\mathcal{H}_{ad}[s_m ] = [1-s_m]\mathcal{H}_{0}+s_m\mathcal{H}(\lambda, \gamma)$ 
for a time $\delta$. The evolution operator for the $m$th step is given by
\begin{equation}
U_{m}  =  e^{-i\delta \mathcal{H}_{ad}[s_m]},
\end{equation}
where $\delta = T_P / (M_P+1)$. The total evolution is
\begin{equation}
U_{P} = \prod _{m=0}^{M_P} U_{m}.
\end{equation}
Since $\mathcal{H}_0$ and $\mathcal{H}(\lambda, \gamma)$ in $\mathcal{H}_{ad}[s_m]$ do not commute, the operator $U_{m}(\delta)$ is approximately implemented by the use of Trotter's formula \cite{Trotter}:
\begin{eqnarray}
U_{m} & = & e^{-i\delta \mathcal{H}_{ad}[s_m]} = e^{-i\frac{\delta}{2}[1-s_m]
\mathcal{H}_{0}} \nonumber \\
&  &   \times e^{-i\delta s_m \mathcal {H}(\lambda,
\gamma)} e^{-i\frac{\delta}{2}[1-s_m] \mathcal {H}_{0}} +
O(\delta^3).
\end{eqnarray}

\begin{figure}
  \includegraphics[width=0.99\columnwidth]{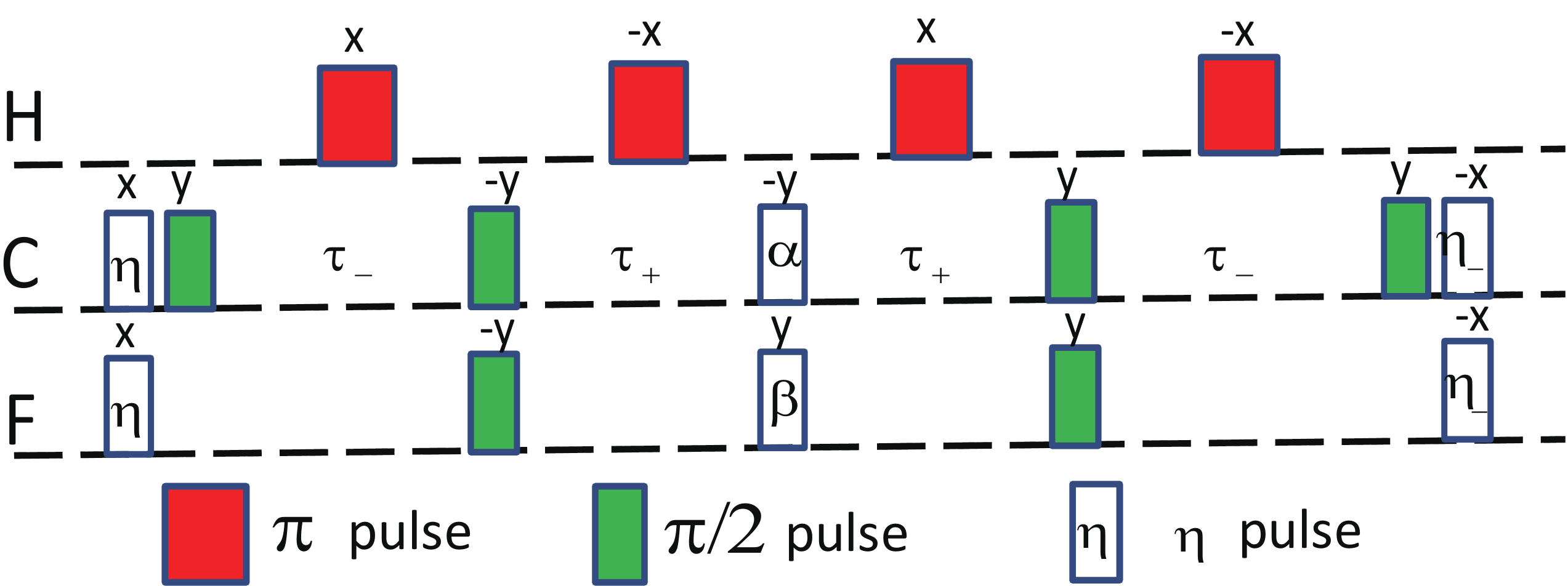}\\
  \caption{The pulse sequence for a step of adiabatic state preparation. Here, $\tau_{\pm}=|\frac{\theta \pm \frac{\pi}{4}}{\pi J_{12}}|$,
  $\eta=(1-s_m)\delta$, $\eta_-=\frac{\pi}{2}-\eta$, $\alpha=(-r-1)s_m\delta$, $\beta=(-r+1)s_m\delta$.} \label{ASPpulse}
\end{figure}

For this stepwise approximation, the duration of each time step $\delta$ has to be
chosen such that (\textit{i}) the time $\delta$ is short enough that
the Hamiltonian simulation holds and (\textit{ii}) the adiabaticity
criterion remains valid, i.e., the total time $T_P$
is long enough. The experimental values $s_m$ for the discretized
scan are represented by black dots in FIG. 2(b) in the paper, which keeps a high
theoretical fidelity (more than 0.99) of the final state for this stepwise transfer process of our ASP.
The two operations $e^{-i\frac{\delta}{2}[1-s_m ]
\mathcal{H}_{0}}$ and $e^{-i\delta s_m \mathcal {H}(\lambda,
\gamma)}$ can be precisely simulated: $e^{-i\frac{\delta}{2}[1-s_m ]
\mathcal{H}_{0}}$ can be easily 
realized using NMR radiofrequency pulses, while we use quantum techniques to simulate the XY Hamiltonian:
$$e^{-i\delta s_m \mathcal {H}(\lambda,
\gamma)}=V^{\dagger}_{d} e^{-i\delta s_m
\mathcal{H}_{d}(\lambda,\gamma)} V_{d},
$$
where the diagonal matrix $\mathcal{H}_{d}(\lambda,\gamma) = V_{d}
\mathcal{H}(\lambda,\gamma) V^{\dagger}_{d}$:
\begin{eqnarray}
\mathcal{H}_{d}(\lambda,\gamma) & = &
\left(
\begin{array}{cccc}
-\sqrt{\lambda^2 + \gamma^2} &  0 & 0 &  0  \\
  0 & -1  & 0 & 0  \\
  0 & 0  & 1 & 0 \\
0 &0   &0  & \sqrt{\lambda^2 + \gamma^2}
\end{array}
\right) \nonumber  \\
& = & - \frac{\sqrt{\lambda^2 + \gamma^2}+1}{2}\sigma_{z}^{1}-\frac{\sqrt{\lambda^2 + \gamma^2}-1}{2}\sigma_{z}^{2} 
\label{Hd}
\end{eqnarray}
with
\begin{eqnarray}
V_{d} & = &
\left(
\begin{array}{cccc}
  \cos(\theta /2) &  0 & 0 &  -\sin(\theta /2)  \\
  0 & 1/\sqrt{2}  & -1/\sqrt{2} & 0  \\
  0 & 1/\sqrt{2}  & 1/\sqrt{2} & 0 \\
\sin(\theta /2)  &0   &0  &   \cos(\theta /2)
\end{array}
\right)  \nonumber  \\
& = &  e^{-i \frac{\theta +\pi/2}{4} \sigma_y^1\sigma_x^2}e^{-i \frac{\theta -\pi/2}{4} \sigma_x^1\sigma_y^2}.
\label{Vd}
\end{eqnarray}
Here $\tan\theta = \gamma/ \lambda$.
Thus the operator $U_m $ can be implemented using a multi-pulse sequence \cite{Peng10}, shown in Fig. \ref{ASPpulse}  . In the case of the unsolved model for the target Hamiltonian, the propagator still can be obtained by average Hamiltonian theory \cite{Ernstbook}.

\begin{figure}
  \includegraphics[width=0.99\columnwidth]{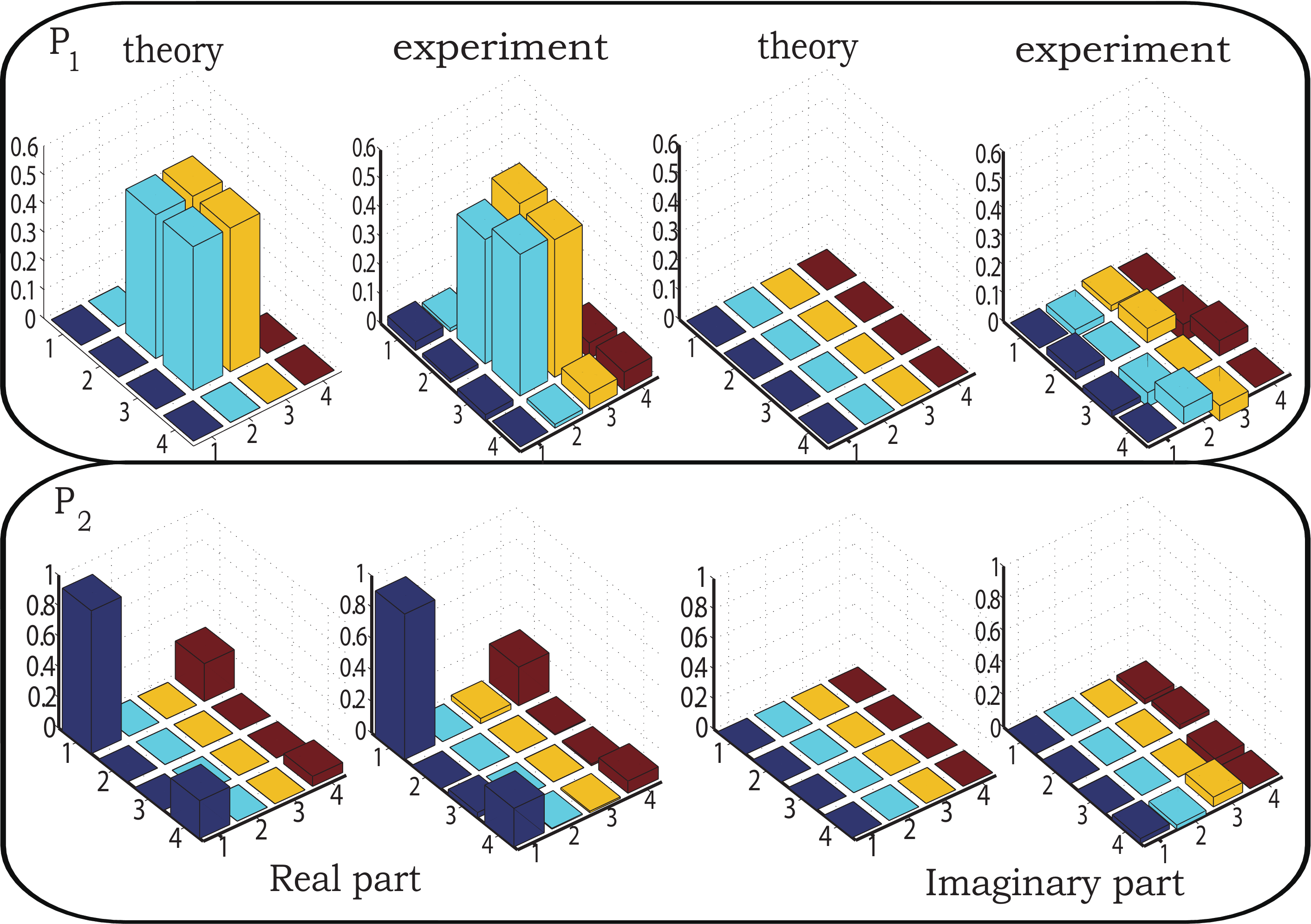}\\
  \caption{Experimental and theoretical deviation density matrices of the system after ASP at the positions $P_1$($\gamma=0.5$,$\lambda=0.878$) and $P_2$($\gamma=0.5$,$\lambda=0$, along with theoretical predictions).
  The left and right columns denote the real and imaginary components, respectively.} \label{tomo}
\end{figure}

To confirm the success of ASP, we performed quantum state tomography on the final state at the end of adiabatic passage. For example, the experimentally reconstructed deviation density matrices of the system at the positions $P_1$ ($\gamma=0.5$,
$\lambda=0$) and $P_2$ ($\gamma=0.5$, $\lambda=0.878$) are shown in Fig. \ref{tomo}, along with the theoretical expectations.
From the tomographically
reconstructed density operators, we determined the experimental fidelity for our prepared states: $F^{exp}_{P_1}=0.98$, $F^{exp}_{P_2}=0.99$. 
This proves that ASP prepared successfully the ground state of the XY model.

\subsection{2. NMR interferometry: A purely GP generation by conditionally cyclic adiabatic variations of the Hamiltonians}

\subsubsection{Cyclic adiabatic evolution}

The studied physical system is described by $ \tilde{\mathcal{H}}(\lambda,\gamma,\phi) =
U_{z}^{\dagger}(\phi) \mathcal{H}(\lambda,\gamma)U_{z}(\phi) $
with $U_{z}(\phi)=\prod_k e^{-i\frac{\phi}{2}\sigma_{z}^{k}}$, which has the same spectrum as $\mathcal{H}(\lambda,\gamma)$,
independent of $\phi$. Its matrix form is 

\begin{equation}
 \tilde{\mathcal{H}}(\lambda,\gamma,\phi)  = - \left(
  \begin{array}{cccc}
    \lambda & 0 & 0 & \gamma e^{i2\phi} \\
    0 & 0 & 1 & 0 \\
    0 & 1 & 0 & 0 \\
    \gamma e^{-i2\phi} & 0 & 0 & -\lambda \\
  \end{array}
\right).
\end{equation}
Clearly we can consider the subspace spanned by $|00\rangle$ and $|11\rangle$,
which contains the ground state, as a pseudo-spin 1/2, where the spin
evolves in an effective magnetic field
$\mathbf{B} = r (\sin \theta \cos 2\phi, \sin \theta \sin 2\phi, \cos \theta) $ with $\tan \theta = \gamma/\lambda$, while the subspace spanned by $|01\rangle$ and $|10\rangle$ is independent of $\phi$. Here, we adiabatically vary the angle $\phi$, that is, the Hamiltonian depends on time through a set of parameters $\mathbf{R}(t) = \mathbf{R}(\theta, \phi(t)) = (\sin \theta \cos 2\phi(t) , \sin \theta \sin 2\phi(t), \cos \theta)$.  
Thus we are interested in the adiabatic evolution of the system as $\mathbf{R}(t)$ moves slowly along a path $C$ in the parameter
space, i.e., 
$$
C: [0,T] \rightarrow \mathcal{S}^2 \mbox{ with points }  \mathbf{R}(t) \in C,
$$
where the unit sphere $\mathcal{S}^2 = \{ \mathbf{R} \in \mathcal{R}^3: | \mathbf{R}| = 1\}$ is the parameter space of the system. 
The quantum adiabatic theorem predicts that a system initially in one of its eigenstates $\vert n;\mathbf{R}(0) \rangle$ will remain its instantaneous eigenstate $\vert n;\mathbf{R}(t) \rangle$ of the Hamiltonian $\tilde{\mathcal{H}}(\mathbf{R}(t))$ throughout the process. 

If the parameters $\mathbf{R}(t)$ adiabatically traverse a closed path $C: [\phi(0) =0, \phi(T) = \pi]$ and return, after some period $T$, to their original values:
$$
C:  \mathbf{R}(0) \rightarrow \mathbf{R}(t) \rightarrow \mathbf{R}(T) =  \mathbf{R}(0), 
$$
then
\begin{eqnarray}
\tilde{\mathcal{H}}(\mathbf{R}(T) ) =  \tilde{\mathcal{H}}(\mathbf{R}(0) ) \nonumber \\
E_n(\mathbf{R}(T) ) = E_n(\mathbf{R}(0) ) \nonumber \\
\vert n;\mathbf{R}(T)  \rangle \langle n;\mathbf{R}(T) \vert =  \vert n;\mathbf{R}(0) \rangle \langle n;\mathbf{R}(0) \vert . \nonumber 
\end{eqnarray}
Here the Hamiltonian $\tilde{\mathcal{H}}(\mathbf{R})$ has the spectral resolution
$$
\tilde{\mathcal{H}}(\mathbf{R}(t)) = \sum_n E_n(\mathbf{R}(t)) \vert n;\mathbf{R}(t)  \rangle \langle n;\mathbf{R}(t) \vert.
$$
However, the basis vectors $\vert n;\mathbf{R}  \rangle$ themselves in general not be unique over the whole parameter space. 
A new set of eigenvectors $\vert n;\mathbf{R}  \rangle'$ can be obtained by gauge transformations: 
$$
\vert n;\mathbf{R}  \rangle' = e^{i \zeta_n(\mathbf{R})}\vert n;\mathbf{R}  \rangle,
$$
where $\zeta_n(\mathbf{R})$ are arbitrary real phase angles. One can use different parameterizations over different patches of the parameter space. Here we require that the closed path $C$ is placed into one single patch $O \subset M$ and the basis functions $ \vert n;\mathbf{R}  \rangle$ is smooth and single-valued.
 
 Consequently, the cyclic adiabatic evolution along the closed path $C: [\phi(0) =0, \phi(T) = \pi]$ let the Hamiltonian  $\tilde{\mathcal{H}}(\mathbf{R}(t))$ and the adiabatically evolving state $\vert \psi(t) \rangle \langle \psi(t) \vert$ return to their original forms in the parameter space as time progresses from $t=0$ to the period $t = T$. If our system start with the ground state $ \vert \Psi_g(0) \rangle = \cos\frac{\theta}{2}\vert 00\rangle + \sin\frac{\theta}{2} \vert 11\rangle$ of $\tilde{\mathcal{H}}(\mathbf{R}(0))$, the adiabatically evolving state at time $t$ is  
 $$
 \vert \psi(t) \rangle = e^{i (\beta_g + \beta_d)}  \vert \Psi_g(\phi) \rangle = e^{i (\beta_g + \beta_d)} U_z^{\dagger}(\phi) \vert \Psi_g(0) \rangle.
 $$
The cyclic adiabatic evolution along the closed path $C$ will generate the dynamical phase
 $$
  \beta_d  = \oint_{0}^{T}E_g(t) dt = \oint_{0}^{T}- \sqrt{\lambda^2 + \gamma^2} dt = - T \sqrt{\lambda^2 + \gamma^2}
 $$ 
 and the geometric phase (or Berry phase)
$$
  \beta_g  = i \oint_{0}^{\pi} \langle \Psi_g(\phi) \vert \partial_{\phi}\vert \Psi_g(\phi) \rangle  = \pi (1- \cos \theta).
 $$
 Berry phase for the GP obtained in the adiabatic approximation is associated with a closed curve in the Hamiltonian parameter space \cite{Berry84}. 
After  the cyclic adiabatic evolution along the closed path $C$, the system state obtains the total phase $\beta_t(C) = \beta_d(C)  + \beta_g(C) = \pi (1- \cos \theta)- T \sqrt{\lambda^2 + \gamma^2}$, mixing the GP with the dynamical phase. 
 
\subsubsection{A purely GP generation: eliminating the dynamical phase}

In order to obtain a purely GP, we have to design a reverse
dynamical process to cancel out the dynamical phase but double the geometric one \cite{GPQC00,GPQC07}. This can be implemented by another closed path $\bar{C}$ ($\phi: \pi \rightarrow 2\pi$) along the Hamiltonian $-\tilde{\mathcal{H}}(\lambda,\gamma,\phi)$. In this case, the initial state $\vert \Psi_g \rangle = \cos\frac{\theta}{2}\vert 00\rangle + \sin\frac{\theta}{2} \vert 11\rangle$ is not the ground state of $-\tilde{\mathcal{H}}(\lambda,\gamma,\phi)$, but the eigenstate of the Hamiltonian $-\tilde{\mathcal{H}}(\lambda,\gamma,\phi)$ with the highest eigenvalue $\sqrt{\lambda^2 + \gamma^2}$. Therefore the adiabatic passage was performed on this eigenstate along $-\tilde{\mathcal{H}}(\lambda,\gamma,\phi)$. Thus we have the dynamical phase
$$
\beta_d(\bar{C}) =  \oint_{0}^{T} \sqrt{\lambda^2 + \gamma^2} dt =  T \sqrt{\lambda^2 + \gamma^2},
$$ 
and the geometric phase
$$
\beta_g(\bar{C}) =  i \oint_{0}^{\pi} \langle \Psi_g(\phi) \vert \partial_{\phi}\vert \Psi_g(\phi) \rangle = \pi (1- \cos \theta),
$$ 
which leads to the total phase $\beta_t(\bar{C}) = \beta_d(\bar{C})  + \beta_g(\bar{C})= \pi (1- \cos \theta) + T \sqrt{\lambda^2 + \gamma^2}$. 

The resulting effect of the two closed paths $C$ and $\bar{C}$ is
$$
\beta_t(\bar{C})  + \beta_t(\bar{C}) = 2 \beta_g = 2\pi (1- \cos \theta).
$$
Consequently, the dynamic phase vanishes and we obtain a purely GP.

\subsubsection{NMR interferometry} 
The phases generated can be detected by NMR interferometry, which consists of a Hadamard gate and a controlled-$U_i$ operation (see Fig. 2 (a) in the paper). The Hadamard gate is represented by the Hadamard matrix:
$$
H = \frac{1}{\sqrt{2}}
\left(
\begin{array}{cc}
1  & 1   \\
 1 &  -1
\end{array}
\right),
$$
which maps the basis state $\vert 0 \rangle$ to $\frac{1}{\sqrt{2}}(\vert 0 \rangle + \vert 1 \rangle)$ and $\vert 1 \rangle$ to $\frac{1}{\sqrt{2}}(\vert 0 \rangle - \vert 1 \rangle)$. 
The auxiliary qubit $a$ was put into a
superposition state $\frac{1}{\sqrt{2}}(\vert 0 \rangle + \vert 1 \rangle)$ from the $\vert 0 \rangle$ state by a pseudo-Hadamard gate $ [\pi /2]_{-y }$.

Then the system state adiabatically traces out a closed path $C$ ($\phi: 0 \rightarrow \pi$) along the Hamiltonian $\tilde{\mathcal{H}}(\lambda,\gamma,\phi) $, but only if the auxiliary qubit is in
state $\vert 1 \rangle$; when the auxiliary qubit is in state $\vert 0 \rangle$, the system state is not affected. This can be realized by a controlled-$U_C$ operation: $\mathcal{U}_C = \vert 0 \rangle \langle 0 \vert_a \otimes \mathbf{1} +  \vert 1 \rangle \langle 1 \vert_a \otimes U_C$,
where $\mathbf{1}$ represents a $4\times 4$ unit operator and the unitary operator $U_C$
is the cyclic adiabatic evolution on the system qubits along the chosen path.
It effectively introduces a relative phase
shift between the initially prepared superposition with
known phase of the states of the auxiliary qubit when the cyclic adiabatic evolution $U_C$ creates a non-zero phase.
The process of the interferometer can be described as
\begin{eqnarray}
\vert 0 \rangle_a \vert \Psi_g \rangle_{12} & \stackrel{H_a}{\longrightarrow} & \frac{1}{\sqrt{2}}(\vert 0 \rangle_a + \vert 1 \rangle_a) \vert \Psi_g \rangle_{12}   \nonumber \\
& \stackrel{\mathcal{U}_{C}}{\longrightarrow} & \frac{1}{\sqrt{2}}(\vert 0 \rangle_a + e^{i [\beta_g(C)  + \beta_d(C) ]}\vert 1 \rangle_a) \vert \Psi_g \rangle_{12} \nonumber.
\end{eqnarray}\label{UC}
Likewise, for the closed path $\bar{C}$, The process is 
\begin{eqnarray}
\vert 0 \rangle_a \vert \Psi_g \rangle_{12} & \stackrel{H_a}{\longrightarrow} & \frac{1}{\sqrt{2}}(\vert 0 \rangle_a + \vert 1 \rangle_a) \vert \Psi_g \rangle_{12}   \nonumber \\
& \stackrel{\mathcal{U}_{\bar{C}}}{\longrightarrow} & \frac{1}{\sqrt{2}}(\vert 0 \rangle_a + e^{i [\beta_g(\bar{C}) + \beta_d(\bar{C})]}\vert 1 \rangle_a) \vert \Psi_g \rangle_{12}.
 \nonumber
\end{eqnarray}\label{UCbar}
The resulting effect of these two experiments results in 
\begin{eqnarray}
\vert 0 \rangle_a \vert \Psi_g \rangle_{12} & \longrightarrow & \frac{1}{\sqrt{2}}(\vert 0 \rangle_a + e^{i 2\beta_g }\vert 1 \rangle_a) \vert \Psi_g \rangle_{12}.
 \nonumber
\end{eqnarray}\label{UCtot}
where the purely GP is obtained by summing the relative phases in these two experiments. 

\subsubsection{Experimental implementation for conditionally cyclic adiabatic evolutions along $C$ and $\bar{C}$}

The Hamiltonian $\tilde{\mathcal{H}}(\lambda,\gamma,\phi)$ varies adiabatically along the trajectory $C$, i.e., $\phi$ changes slowly from 0 to $\pi$.
Like in APS, the continuous Hamiltonian $\tilde{\mathcal{H}}(\lambda,\gamma,\phi)$ is discretized into $M + 1$ steps in the range of $\phi: 0 \rightarrow \pi$ in the actual implementation. Likewise, we numerically optimized the adiabatic steps $M + 1$ and evolution time $T$ to achieve a high fidelity of the instantaneous state of the system. In experiment, we chose $M<6$ and $T \sim 10$ which results
in a theoretical fidelity of $>0.99$ for both trajectories $C$ and $\bar{C}$.

\begin{figure}
  \includegraphics[width=0.99\columnwidth]{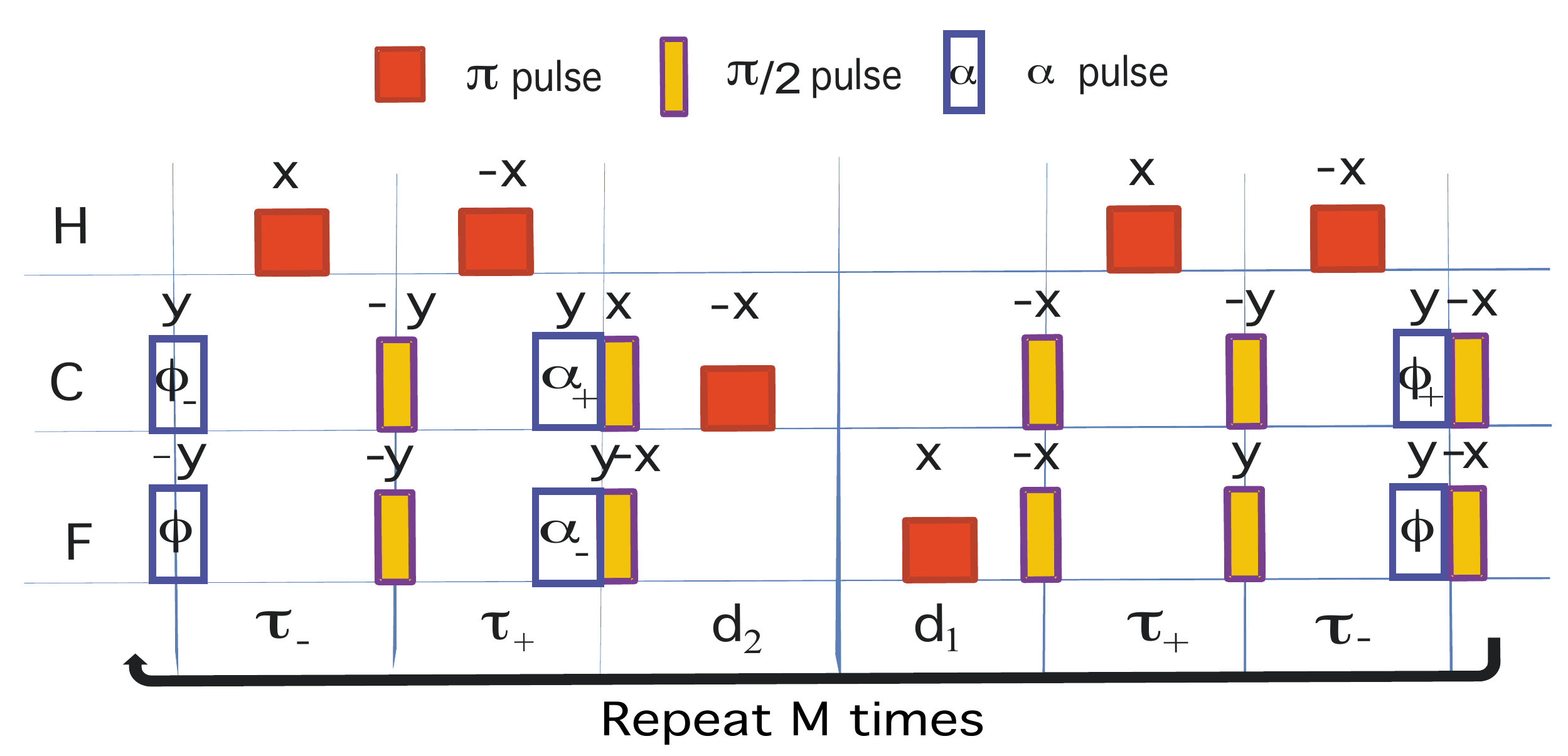}\\
  \caption{Pulse sequence for implementing the control operation $\mathcal{U}_C$
  for the adiabatic path $C$.
  Here $\tau_{\pm}=|\frac{\theta \pm \frac{\pi}{4}}{\pi J_{12}}|$, $d_{i}=|\frac{T(1 - (-1)^i r)}{2M\pi J_{ai}}|$ $(i=1,2)$,
$\phi_{\pm}=\frac{\pi}{2}\pm \phi$,  and $\alpha_{\pm}=\frac{T(r \pm 1)}{2M}$.}
\label{Pulseq}
\end{figure}

The unitary operation of the $m$th adiabatic step for a constant $\phi_m$ can be realized by the following decomposition:
\begin{eqnarray}
\mathcal{U}^m_{C} & = &
e^{-i\frac{1}{2}(\mathbf{1}^{a}-\sigma_{z}^{a})\otimes
\tilde{\mathcal{H}}(\lambda,\gamma,\phi_m) \tau} \nonumber\\
&=& \mathbf{1}^{a} \otimes U_{z}^{\dagger}(\phi_m) e^{-i\frac{1}{2}(\mathbf{1}^{a}-\sigma_{z}^{a}) \otimes
\mathcal{H}(\lambda,\gamma) \tau }\mathbf{1}^{a} \otimes U_{z}(\phi_m) \nonumber \\
 &=&\mathbf{1}^{a} \otimes U_{z}^{\dagger}(\phi_m)  V^{\dagger}_{d}e^{-i\frac{1}{2}(\mathbf{1}^{a}-\sigma_{z}^{a})\otimes
\mathcal{H}_{d}(\lambda,\gamma) \tau}\mathbf{1}^{a} \otimes
V_{d} U_{z}(\phi_m), \nonumber
\end{eqnarray}\label{UC}
where $\tau = T/(M+1)$. The total evolution is
\begin{equation}
U_{C}(T)  = \prod _{m=0}^{M} \mathcal{U}^m_{C}.
\end{equation}
$U_{z}^{\dagger}(\phi_m)$ is a rotation  of the system qubits around the $z$ axis, which can be realized by $\prod_k R_{kx}(\pi/2)R_{ky}(\phi_m)R_{kx}(-\pi/2)$ 
and $V^{\dagger}_{d}$ by Eq. (\ref{Vd}).
The conditional operation
\begin{eqnarray}
e^{-i\frac{1}{2}(\mathbf{1}^{a}-\sigma_{z}^{a}) \otimes
\mathcal{H}_d(\lambda,\gamma) \tau } \nonumber \\
= e^{i\frac{1}{2}(\mathbf{1}^{a}-\sigma_{z}^{a})\otimes
(\frac{r+1}{2}\sigma_{z}^{1}+\frac{r-1}{2}\sigma_{z}^{2})
\tau}  \nonumber \\
= e^{i(\frac{r+1}{4}\sigma_{z}^{1}+\frac{r-1}{4}\sigma_{z}^{2})\tau} e^{-i\frac{r+1}{2}\sigma_{z}^{a}\sigma_{z}^{1} \tau}e^{-i\frac{r-1}{2}\sigma_{z}^{a}\sigma_{z}^{2} \tau} \nonumber
\end{eqnarray}
is also implemented by rf pulses and J-coupling evolutions, shown in Fig. \ref{Pulseq}. 

For the trajectory $\bar{C}$, the propagator is
\begin{eqnarray}
\mathcal{U}^m_{\bar{C}} & = &
e^{-i\frac{1}{2}(\mathbf{1}^{a}-\sigma_{z}^{a})\otimes
[- \tilde{\mathcal{H}}(\lambda,\gamma,\phi_m)] \tau} \nonumber\\
& = &
e^{-i\frac{1}{2}(\mathbf{1}^{a}-\sigma_{z}^{a})\otimes [
R_{lz}^{\dagger}(\pi) \tilde{\mathcal{H}}
(-\lambda, \gamma,\phi_m)R_{lz}(\pi) ] \tau} \nonumber\\
&=& \mathbf{1}^{a} \otimes U_{z}^{\dagger}(\phi_m) R_{k\mu}(\pi) R_{j\nu}(\pi)  e^{-i\frac{1}{2}(\mathbf{1}^{a}-\sigma_{z}^{a}) \otimes
\mathcal{H}(\lambda,\gamma)\tau} \nonumber \\
& & \times \mathbf{1}^{a} \otimes R_{k\mu}(\pi) R_{j\nu}(\pi) U_{z}(\phi_m) \nonumber \\
 &=&\mathbf{1}^{a} \otimes R_{k\mu}(\pi) R_{j\nu}(\pi)  U_{z}^{\dagger}(- \phi_m)  V^{\dagger}_{d}e^{-i\frac{1}{2}(\mathbf{1}^{a}-\sigma_{z}^{a})\otimes
\mathcal{H}_{d}(\lambda,\gamma) \tau} \nonumber \\
& & \times \mathbf{1}^{a} \otimes
V_{d} U_{z}(-\phi_m) R_{k\mu}(\pi) R_{j\nu}(\pi). \nonumber
\end{eqnarray}\label{UCbar}
with $k,j = 1 \mbox{ or } 2, (k \neq j) $, and $\mu,\nu = x \mbox{ or } y, (\mu \neq \nu)$.

\subsection{3. Phase measurement: Quadrature detection in NMR}

\begin{figure}
  \includegraphics[width=0.99\columnwidth]{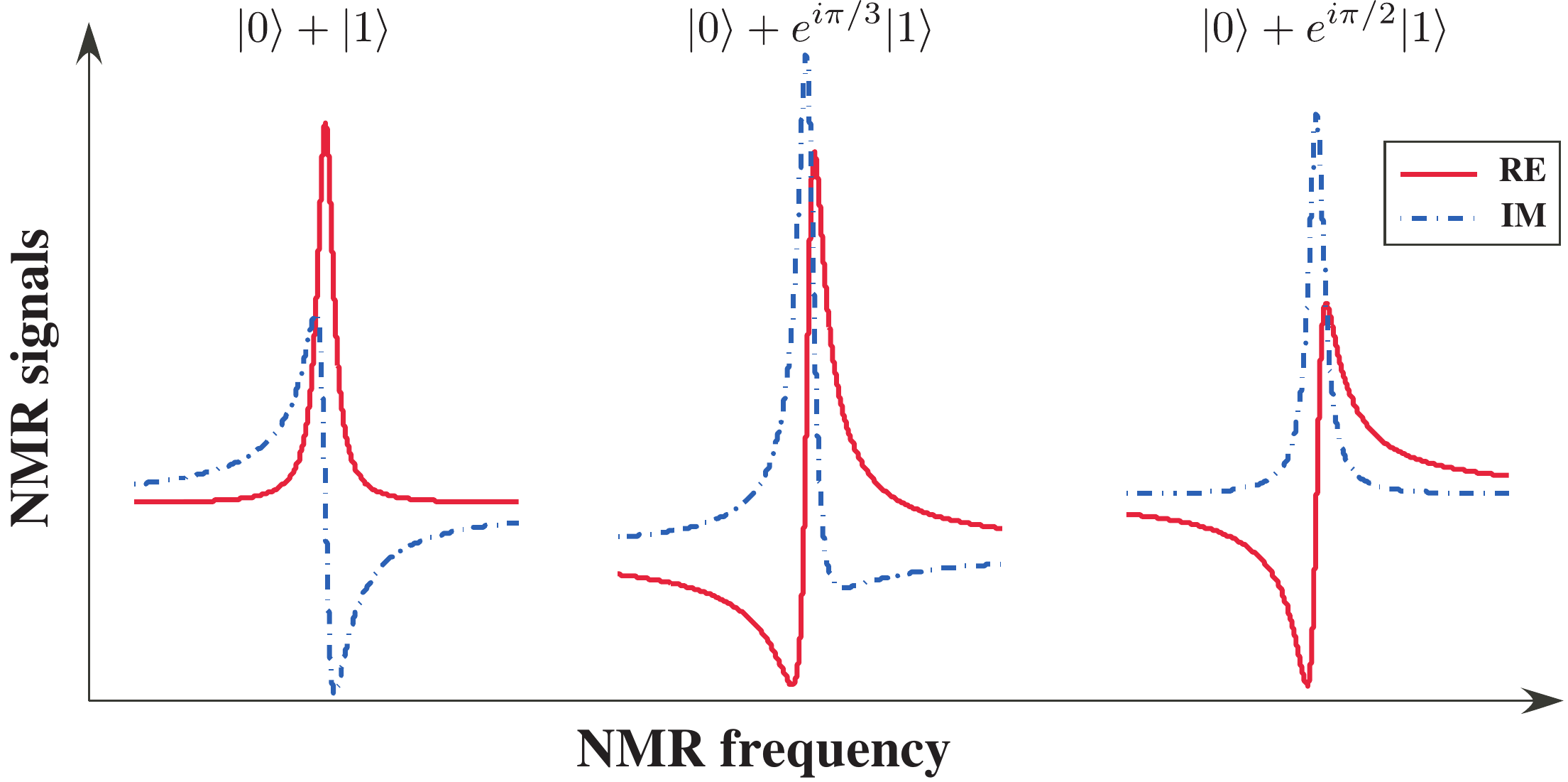}\\
  \caption{Phase measurement by quadrature detection in NMR.} \label{phasemeas}
\end{figure}

After NMR interferometry, the state of the quantum register is 
$\frac{1}{\sqrt{2}}(\vert 0 \rangle_a + e^{i (\beta_g + \beta_d)}\vert 1 \rangle_a) \vert \Psi_g \rangle_{12}$.  
A relative phase shift $\beta_g + \beta_d$ is created between the states $\vert 0 \rangle_a$ and $\vert 1 \rangle_a$ of the auxiliary  qubit.  It can be obtained when we measure
the NMR signal of the auxiliary qubit:
$$ \langle \sigma^{-}_a
\rangle =\frac{1}{2} \langle\sigma_{x}^{a}-i\sigma_{y}^{a}\rangle =
\frac{1}{2}[\cos (\beta_g + \beta_d) + i \sin (\beta_g + \beta_d)].
$$
The quadrature detection in NMR serves as a phase sensitive demodulation technique, \emph{i.e.}, the complex demodulated signal is separated into two components (the real part $RE \propto \cos (\beta_g + \beta_d)$ and the imaginary part $IM \propto \sin (\beta_g + \beta_d)$) which are 90¡ out of phase with each other. Thus, the phase angle $\beta_g + \beta_d$ of the signal can be determined by $\tan^{-1} (IM/RE)$. Quadrature detection combined with Fourier analysis thus gives all the necessary information on the magnetic resonance signal components i.e., amplitude, phase and frequency \cite{Ernstbook}. Taking the input state of $\frac{1}{\sqrt{2}}(\vert 0 \rangle_a +\vert 1 \rangle_a) \vert \Psi_g \rangle_{12}$ as the reference spectrum, we measured the relative phase information by the phase of the Fourier-transformed spectrum. Fig. \ref{phasemeas} shows a simple example for the phase measurement by the Fourier-transformed spectra.

\subsection{Experimental results and data analysis}

\begin{figure}
  \includegraphics[width=0.99\columnwidth]{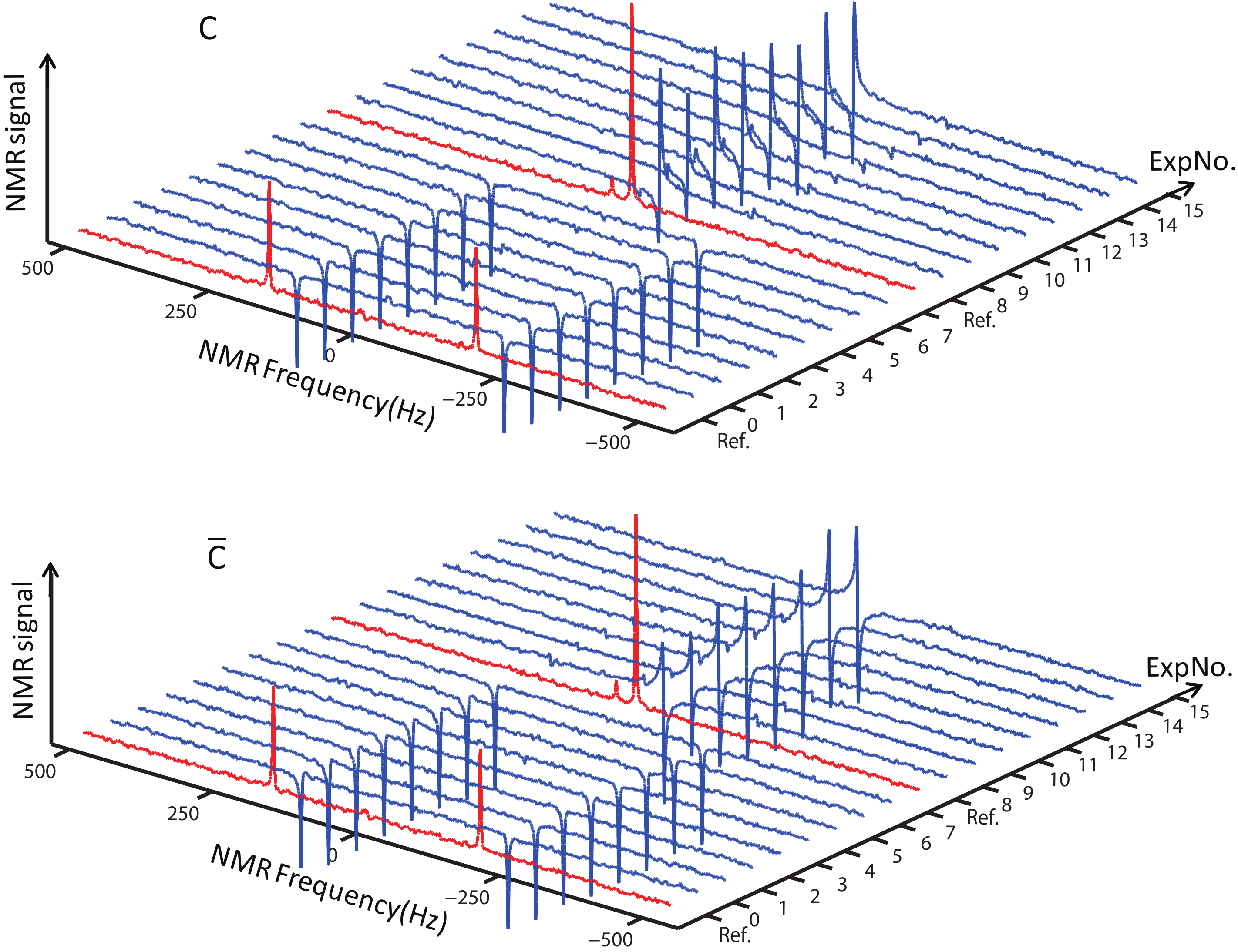}\\
  \caption{Experimental $^{13}$C spectra for the XY Hamiltonian $\mathcal{H}(\lambda,\gamma)$ with $\gamma = 0.5$ and the variable $\lambda$ from 0 to 1.7. The red ones are the spectra of the initial states prepared by ASP as the references. The upper plot is the spectra for the trajectory $C$ when $\lambda$ scans while the lower plot is for $\bar{C}$. We extract the phase information from these spectra using quadrature detection.} \label{ExpSpetra}
\end{figure}

Fig. \ref{ExpSpetra} shows the experimental NMR spectra for a set of experiments $\mathcal{H}(\lambda,\gamma)$ with
varying magnetic field strength $\lambda$ $(0 \rightarrow 1.7)$ and a constant anisotropy parameter
$\gamma = 0.5$. The parameter $\lambda$ was varied by a hyperbolic sine function \cite{Peng05PRA}, but avoiding the level crossing points.
From these spectra, we measured the phase shifts accumulated by the two trajectories $C$ and $\bar{C}$ listed in Table I. 
The pure GP was then obtained by summing the two phase shifts from $C$ and $\bar{C}$: $\beta_g = (\beta_t(C) + \beta_t(\bar{C}))/2$.
A set of the experimental spectra are shown in Fig. 3 in the paper.

\begin{table}
\caption{The extracted phase values from the NMR signals for
$\gamma=0.5$. $\beta_{C}^{exp}$ and $\beta_{\bar{C}}^{exp}$ denote
the phases obtained by the NMR interferometry with the related
adiabatic evolution trajectories $C$ and $\bar{C}$. $\beta_{g}^{exp}
=\frac{\beta_{C}^{exp}+\beta_{\bar{C}}^{exp}}{2}$ is the geometric
phase. The corresponding experiment spectra are showed
above.}\label{expdata}
\begin{tabular}[t]{|c|c|c|c|c|c|}
 \hline
ExpNo. & $\lambda$ &  $\beta_{C}^{exp}$ ($^{\circ}$) & $\beta_{\bar{C}}^{exp}$($^{\circ}$) & $\beta_{g}^{exp}$ ($^{\circ}$)& $\beta_{g}^{th}$ ($^{\circ}$) \\
\hline

    0 & 0 &170.6 & -173 & -1.2 &0 \\
  1 &  0.327 &174 & -171.4 & 1.3&0 \\
  2 & 0.5312 & 173.4 & -172 & 0.7&0 \\
  3 &  0.6592  &173.8 & -174.2 & -0.3& 0\\
  4 &  0.74 &168 & -170 & -1 &0\\
  5 & 0.7922& 172.2 & -171 & 0.6 &0\\
  6 & 0.8275 & 174.6 & -176.2 & -0.8&0 \\
  7 &  0.8541 & 170.4 & -171.6 & -0.6 &0\\
  8 &   0.878 &-62.6 & 114.2 & 25.8  & 23.6\\
  9 &   0.9045 & -60 & 107.6 & 23.8 & 22.5\\
  10 &   0.9399 & -60.2 & 106.4 & 23.1 & 21.1\\
  11 &  0.992 & -69.8 & 111.4 & 20.8& 19.3 \\
  12 & 1.0729 & -70.4 & 108.6 & 19.1 & 16.8\\
  13 & 1.2009  & -79.6 & 109 & 14.7 & 13.8 \\
  14 & 1.4051 & -75.6 & 95.4 & 9.9 & 10.4 \\
  15 & 1.7321 & -81.1 & 96.6 & 7.7 & 7.1 \\
\hline
\end{tabular}
\end{table}

Our experimental errors of the geometric phases are less than $3^\circ$. 
These errors result from the imperfection of the initial ground state by ASP,
the diabatic effect, and other experimental imperfections such as the inhomogeneity of the radio frequency field and the
static magnetic field, and the imperfect calibration of the
radio frequency pulses. 
The decoherence from spin relaxation
was small, since the total experimental time of less than 90 $ms$ was short compared to the shortest relaxation
time of 1.0 $s$.

The error contributed by the imperfection of the initial ground state of ASP can be evaluated by the use of the measured input density matrices $\rho_{ini}^{exp}$ after ASP, e.g, shown in Fig. \ref{tomo}. Therefore we have the input state
$$
\rho_{in}=\vert 0 \rangle_a  \langle 0 \vert \otimes \rho_{ini}^{exp}.
$$
Then we input the state to an ideal NMR interferometer, i.e, a
perfectly implemented Hadamard gate and controlled evolution process simulated on a
classical computer to get the theoretical output:
$$
\rho_{f}
=\frac{1}{2}\left(
                       \begin{array}{cc}
                         \rho_{ini}^{exp} & \rho_{ini}^{exp}U^{\dagger}_{\alpha}(T) \\
                         U_{\alpha}(T)\rho_{ini}^{exp} & U_{\alpha}(T)\rho_{ini}^{exp}U^{\dagger}_{\alpha}(T) \\
                       \end{array}
                     \right) ,
  $$
where $\alpha$ ($\alpha=C$ or $\bar{C}$).
As a result, the measurement on the auxiliary qubit by the quadrature detection gives
$$
\langle \rho_{f} \sigma^{-}_a \rangle =  \frac{1}{2}
Tr[U_{\alpha}(T)\rho_{ini}^{exp}] .
$$
Thus we achieved the simulated phases $\beta_{\alpha}^{sim}=
arg(\langle \rho_{f} \sigma^{-}_a \rangle) =
arg[Tr(U_{\alpha}(T)\rho_{ini}^{exp})] $. Thus the simulated
geometric phase starting from the experimental initial state is
$\beta_{g}^{sim}= \frac{\beta_{C}^{sim}+\beta_{\bar{C}}^{sim}}{2}$.
We found that the errors contributed by the imperfection of the
prepared initial state is about $1^\circ$.

\subsection{References}
\begin{itemize}
\bibitem{Trotter} H. F. Trotter, PaciÞc J. Math. \textbf{8}, 887 (1958).
\bibitem{PPS} I. L. Chuang et al., Proc. R. Soc. A \textbf{454}, 447 (1998); D. G. Cory, A. F. Fahmy, and T. F. Havel, Proc. Natl. Acad. Sci.
U.S.A. \textbf{94}, 1634 (1997).
\bibitem{QST} G. M. Leskowitz and L. J. Mueller, Phys. Rev. A \textbf{69}, 052302 (2004).
\bibitem{Peng10}X. Peng and D. Suter, Front. Phys. China \textbf{5}, 1-25 (2010).
\bibitem{Ernstbook}R. R. Ernst, G. Bodenhausen, and A. Wokaun. \textit{Principles of Nuclear Magnetic Resonance in One and
Two Dimensions}. Oxford University Press, Oxford, 1994.
\bibitem{Berry84}M. V. Berry, Proc. R. Soc. Lond. A \textbf{392}, 45 (1984).
\bibitem{Peng05PRA}X. Peng \emph{et al.}, Phys. Rev. A \textbf{71}, 012307 (2005).
\end{itemize}

\end{document}